\documentclass[twocolumn,prl]{revtex4}
\usepackage {graphicx}
\usepackage{subfigure}
\begin{document}

\title{Stripe formation in differentially forced binary systems}
\author{C. M. Pooley and J. M. Yeomans}

\address{The Rudolf Peierls Centre for Theoretical Physics, 1 Keble Road,
Oxford, OX1 3NP, United Kingdom.
}
\date{\today}
\begin{abstract}
We consider pattern formation in periodically forced binary
systems. In particular we focus on systems in which the two species
are differentially forced, one being accelerated with respect to the
other. Using a continuum model consisting of two isothermal ideal
gases which interact via a frictional force we
demonstrate analytically that stripes form spontaneously above a
critical forcing amplitude. The wavelength of
the stripes is found to be close to the wavelength of sound in the
limit of small viscosity. The results are
confirmed numerically. We suggest that the same mechanism
may contribute to the formation of stripes in experiments on
horizontally oscillated granular mixtures.
\end{abstract}
\maketitle

Binary systems subject to an oscillatory driving force are often found
to phase segregate and form patterned structures \cite{SM01, B97}. 
Often this segregation can have important practical applications.
For example it may provide a useful way of separating two mixed
species, or conversely, it may be undesirable in an industrial
process \cite{W}. 
The phenomenon is well documented experimentally but a full
theoretical understanding is still missing with different mechanisms
for the pattern formation being proposed in the literature.

Sanchez {\it et al.} \cite{SS} have carried out experiments and parallel simulations on
granular mixtures immersed in water. For sufficiently large amplitudes
of vibration a mixture of equal sized glass and bronze particles were
found to separate into a striped pattern. Sanchez {\it et al.} proposed that the
differential fluid drag on the two components of the mixture is the mechanism
responsible for the segregation.

Mullin {\it et al.} \cite{M00, RM02, RE03} have performed a series of 
experiments in which they horizontally
oscillate a quasi-2 dimensional layer of bronze spheres and poppy
seeds. The species are again observed to segregate into a striped pattern.
However they propose a different explanation for 
the stripes: the effective excluded volume
interactions which occur because of the different size of the shaken
particles \cite{AO58}.

Therefore our aim in this Letter is to investigate possible mechanisms for
binary phase segregation by providing an approximate analytic solution to a
one-dimensional isothermal continuum model describing the physics of a binary
mixture of two ideal fluids. We find that if the components are
differentially forced stripes are formed  in the concentration above a
critical forcing amplitude. The wavelength of the stripes is
controlled by the velocity of sound and standing wave oscillations in
the total density pay a crucial role in the stripe formation.

We then relate our results more closely to experiments on granular 
mixtures through a simple particle model. We find that differential
forcing of the two types of particle is enough to cause stripe formation
even if the particles have the same mass and volume.

We consider two ideal fluids, labeled $A$ and $B$, which are coupled
by a frictional force proportional to the difference in their
velocities. The physical origin of this force is from collisions 
between the $A$ and $B$ particles which tend to equalize the
velocities. The $A$ fluid is accelerated by a periodic force 
$a \cos(\omega t)$. The Navier-Stokes and continuity equations for
this system may be written \cite{MY99}
\begin{eqnarray}
\rho^A D_t^A u^A \!\!\! &=& \!\!\!-\theta \partial_x \rho^A + \xi (u^B-u^A) + \partial_x (\nu \rho^A \partial_x u^A)\nonumber\\
&& + a \rho^A \cos(\omega t), \label{nss1}\\
\rho^B D_t^B u^B \!\!\!&=& \!\!\!-\theta \partial_x \rho^B + \xi (u^A-u^B) + \partial_x (\nu \rho^B \partial_x u^B), \label{ns2}\\
D_t^A \rho^A \!\!\!&=& \!\!\!-\rho^A \partial_x u^A \label{ce1},\\
D_t^B \rho^B \!\!\!&=&\!\!\! -\rho^B \partial_x u^B \label{ce2},
\label{eqns}
\end{eqnarray}
where $\theta = {k_BT}/{m}$ is the reduced temperature, $\nu$ is the 
kinematic viscosity, and $\rho^A$, $\rho^B$ and $u^A$, $u^B$ are the 
densities and velocities of the $A$ and $B$ fluids. $D_t^A = 
\partial_t + u^A \partial_x$ and $D_t^B = \partial_t + u^B \partial_x$ 
are the material derivatives in the $A$ and $B$ frames, and the
friction coefficient can be written $\xi = \xi^0 {\rho^A
\rho^B}/{\rho}$, where $\xi^0$ is constant and $\rho = \rho^A +
\rho^B$ is the total density of the mixture.
\begin{figure}
\begin{center}
\begin{tabular}{c}
(a)\includegraphics[width = 6cm]{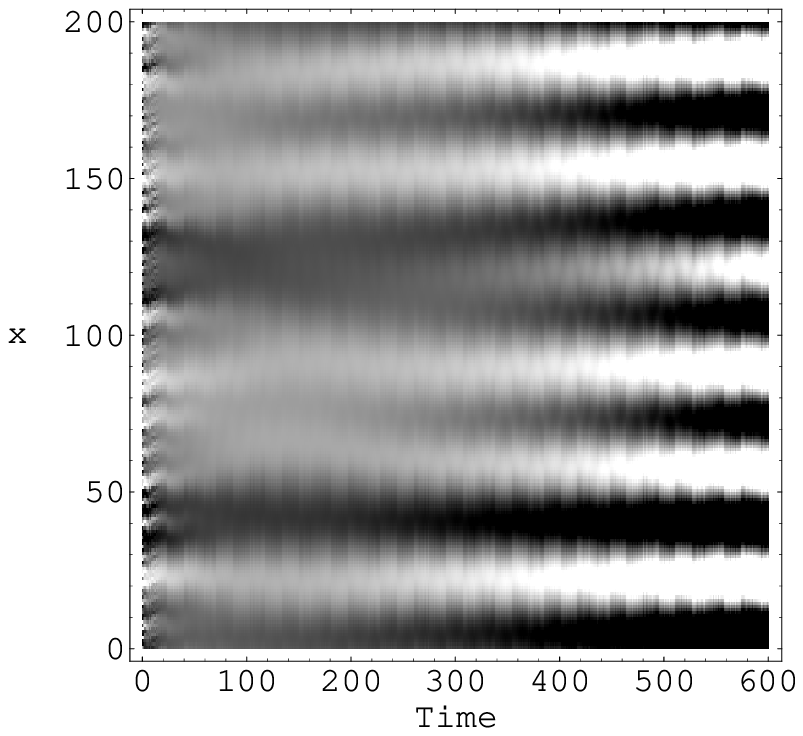}\\
(b)\includegraphics[width = 6cm]{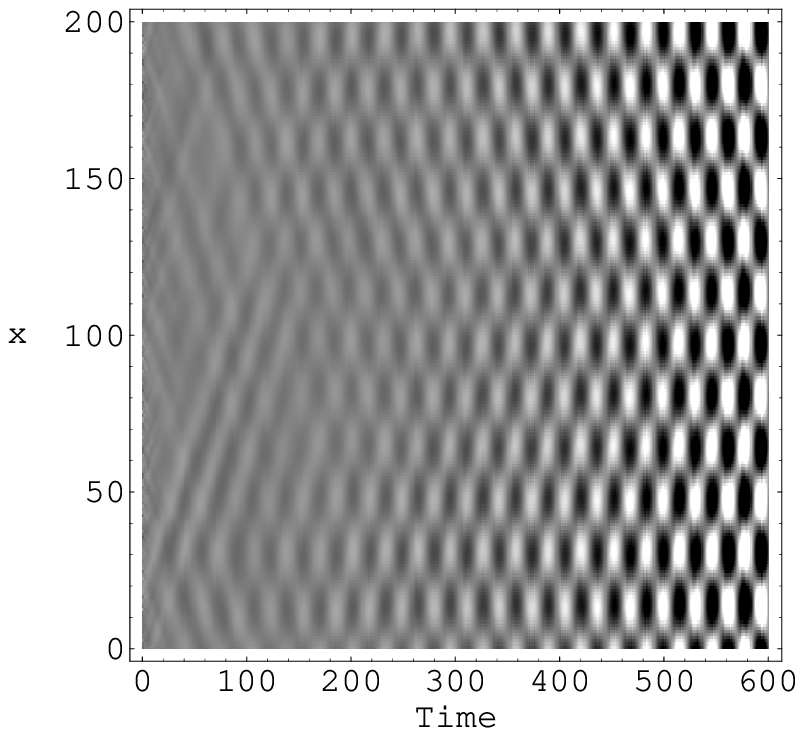}
\end{tabular} 
\end{center}
\caption{Evolution in time of (a) the concentration 
and (b) the total density of a differentially forced binary 
fluid as a function of the distance $x$ along a one dimensional 
system, observed in the rest frame of the fluid.  The concentration
forms stripes and the total density performs standing wave 
oscillations, which increase in amplitude, at the 
frequency of the applied force.
The parameters used were $\xi^0 = 6$, $\nu = 0.5$, $\theta = 1$, 
$a = 0.358$, and $\omega = 0.2$. The wavelength of the 
resulting stripes $\lambda = 33$ is close to the 
wavelength of sound $\frac{2 \pi \sqrt{\theta}}{\omega} = 31$.}
\label{timeev}
\end{figure}

Fig.~\ref{timeev} shows the results of numerically solving 
Eqns.~(\ref{nss1}--\ref{ce2}) using a lattice Boltzmann algorithm. 
Initially the densities of the two components have the same constant 
value $\rho_0$ plus a small random perturbation. Provided the 
amplitude of the forcing $a$ is sufficiently high, strongly segregated
$A$ and $B$ stripes form as shown in Fig. \ref{timeev}a.
These correspond to an approximately sinusoidal modulation of the
concentration $\phi = \rho^A/\rho$. We define the average                             
velocity of the fluid  $u=({\rho^A u^A + \rho^B u^B})/{\rho}$ and the                
rest frame of the fluid to be that moving at velocity $\left< u
\right>$, where $\left< ... \right>$ denotes a spatial average. 
In this frame the total density performs standing wave oscillations 
at the forcing frequency but shifted in phase. 

We assume that the coupling between the two components is sufficiently 
strong that it is the fastest relaxation process in the system. 
Therefore $u^A$ and $u^B$ are always very close in value; hence their 
time derivatives are essentially the same. This allows us to
approximate $D_t^A u^A = D_t^B u^B = D_t u$, 
where $D_t = \partial_t + u \partial_x$. 
Adding equations (\ref{nss1}) and (\ref{ns2}) to remove the coupling
term then gives
\begin{eqnarray}
D_t u = \frac{-\theta}{\rho} \partial_x \rho + a \phi \cos(\omega t) + \nu \partial_x^2 u.
\label{Dtu}
\end{eqnarray}
Similarly adding Eqns.~(\ref{ce1}) and (\ref{ce2})
\begin{equation}
D_t \rho = -\rho \partial_x u.
\label{Dtrho}
\end{equation} 
Motivated by the numerical results we consider the trial solutions
\begin{eqnarray}
\!\!\phi\!\! &=& \!\! \frac{1}{2}\! +\! \Delta \phi \sin(k x),\label{phi} \\
\!\!\rho\!\! &=& \!\! 2 \rho_0\! +\! \cos(k x) \left[ \Delta \rho_1 \cos(\omega t) 
+ \Delta \rho_2 \sin(\omega t) \right],\label{nanb}\\
\!\! u \!\! &=& \!\!\frac{a \sin(\omega t)}{2 \omega}\! + \!\sin(kx) \left[ \Delta u_1
\cos(\omega t) + \Delta u_2 \sin( \omega t) \right],
\label{u}
\end{eqnarray} 
where $k = \frac{2 \pi}{\lambda}$ is the wavenumber of the stripes.

Substitution of the trial solutions (\ref{phi}--\ref{u}) into
Eqns. (\ref{Dtu}) and (\ref{Dtrho}), and comparing cosine and sine terms gives
\begin{eqnarray}
\Delta \rho_1 &=& \left[ \frac{2 \rho_0 ka (\omega^2 - \theta k^2)}
{ (\omega^2 - \theta k^2)^2 + \nu^2 k^4 \omega^2} \right] \Delta \phi,\label{ratiosf}\\
\Delta \rho_2 &=& \left[ \frac{-2 \rho_0 k^3 a \nu \omega}{ (\omega^2 - \theta k^2)^2 + \nu^2 k^4 \omega^2} \right] \Delta \phi,\\
\Delta u_1 &=& \left[ \frac{k^2 a \nu \omega^2}{ (\omega^2 - \theta k^2)^2 + \nu^2 k^4 \omega^2} \right] \Delta \phi,\\
\Delta u_2 &=& \left[ \frac{a \omega (\omega^2 - \theta k^2)}{ (\omega^2 - \theta k^2)^2 + \nu^2 k^4 \omega^2} \right] \Delta \phi.
\label{ratios}
\end{eqnarray}

The crucial information needed is whether a wave with wavenumber $k$ 
grows or decays, and the rate of this process. To understand this we 
must look at the behaviour of $\phi$ as a function of time. The
material derivative must now be treated exactly because the
concentration changes much more slowly than densities and
velocities. Using the continuity equations (\ref{ce1}) and (\ref{ce2})
and $\rho^A \approx \rho^B \approx \rho_0$ gives
\begin{eqnarray}
D_t \phi &=& \frac{1}{4} \partial_x (u^B-u^A) + \frac{u^B-u^A}{8 \rho_0} \partial_x \rho.
\label{dtphi}
\end{eqnarray}
Rearranging Eqn.~(\ref{ns2}) and substituting the trial solutions  
\begin{eqnarray}
u^B - u^A &=& - \frac{a}{\xi^0} \cos(\omega t) 
- \frac{a \omega}{{\xi^0}^2} \sin(\omega t) \nonumber\\
&& + \frac{4 \theta \Delta \phi k}{\xi^0} \cos(kx)+ \frac{2 \zeta}{\xi^0} \sin(kx), 
\label{ubua2}
\end{eqnarray}
where 
\begin{eqnarray}
\!\!\!\! \zeta &=& \left( \omega \Delta u_1 - \nu k^2 \Delta u_2 + 
\frac{\theta k}{2 \rho_0} \Delta \rho_2 \right)  \sin(\omega t)\nonumber\\
 &&\!\!\!\!\!\!+ \left( - \nu k^2 \Delta u_1 - \omega \Delta u_2 + 
\frac{\theta k}{2 \rho_0} \Delta \rho_1 \right) \cos(\omega t).
\end{eqnarray}
Comparing the first order terms in $u_B-u_A$ in Eqn.~(\ref{ubua2}) 
and in $u$ in Eqn.~(\ref{u}) we see that the condition that $A$ and
$B$ are strongly coupled is equivalent to stating $\xi^0 >> \omega$.

Substituting (\ref{nanb}) and (\ref{ubua2}) into (\ref{dtphi}) leads to 
\begin{eqnarray}
D_t \phi &=& \frac{1}{\xi^0} \left( - \theta \Delta \phi k^2 \sin(kx) 
+ \frac{\zeta k}{2} \cos(kx)\right. \nonumber\\
&& \!\!\!\!\!\!\!\!\!\!\! \left.	+ \frac{ka}{8 \rho_0} \cos(\omega t) \sin(kx) \left
[ \Delta \rho_1 \cos(\omega t) 
+ \Delta \rho_2 \sin(\omega t) \right] \right).\nonumber
\label{dtphi3}
\end{eqnarray}
The time scale over which $\phi$ changes is much longer than that
taken for a single oscillation of the force.
This allows us to average the right hand side over time
\begin{eqnarray}
D_t (\Delta \phi) &=& \frac{1}{\xi^0} \left( - \theta k^2 + \frac{k a  \Delta \rho_1 }{16 \rho_0 \Delta \phi} \right) \Delta \phi. 
 \label{dtphi4}
\end{eqnarray}
Substituting expression (\ref{ratiosf}) for $\Delta \rho_1$ and
integrating gives the exponential solution
\begin{eqnarray}
\Delta \phi = e^{\Gamma t},
\label{grow}
\end{eqnarray}
where
\begin{eqnarray}
\Gamma = \frac{1}{\xi^0} \left[ \frac{k^2 a^2}{8} \left( \frac{\omega^2 - \theta k^2}{(\omega^2 - \theta k^2)^2 + \nu^2 k^4 \omega^2} \right) - \theta k^2  \right].
\label{gama}
\end{eqnarray}
\begin{figure}
\begin{center}
\begin{tabular}{c}
(a)\includegraphics[width = 6cm]{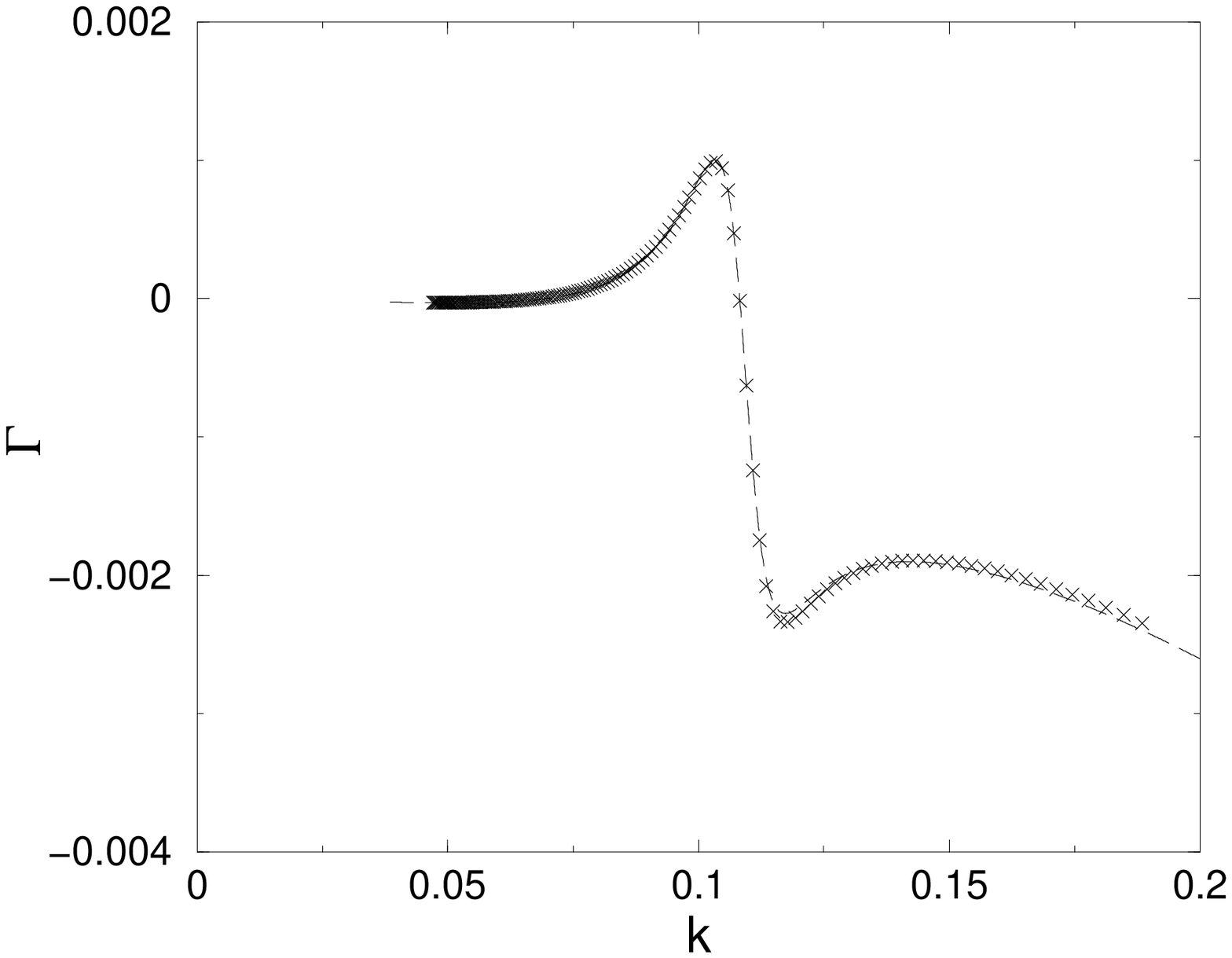}\\
(b)\includegraphics[width = 6cm]{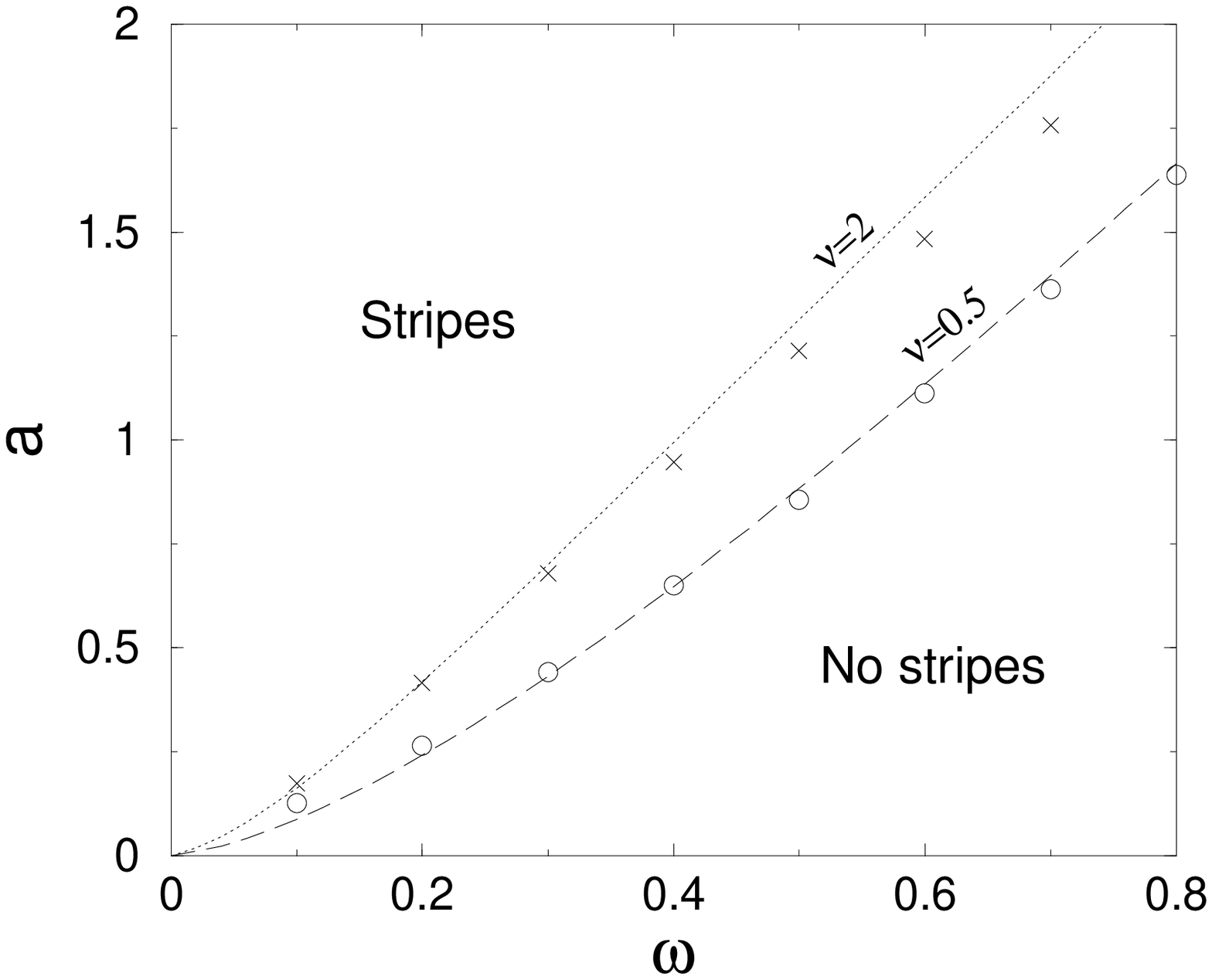}\\
(c)\includegraphics[width = 6cm]{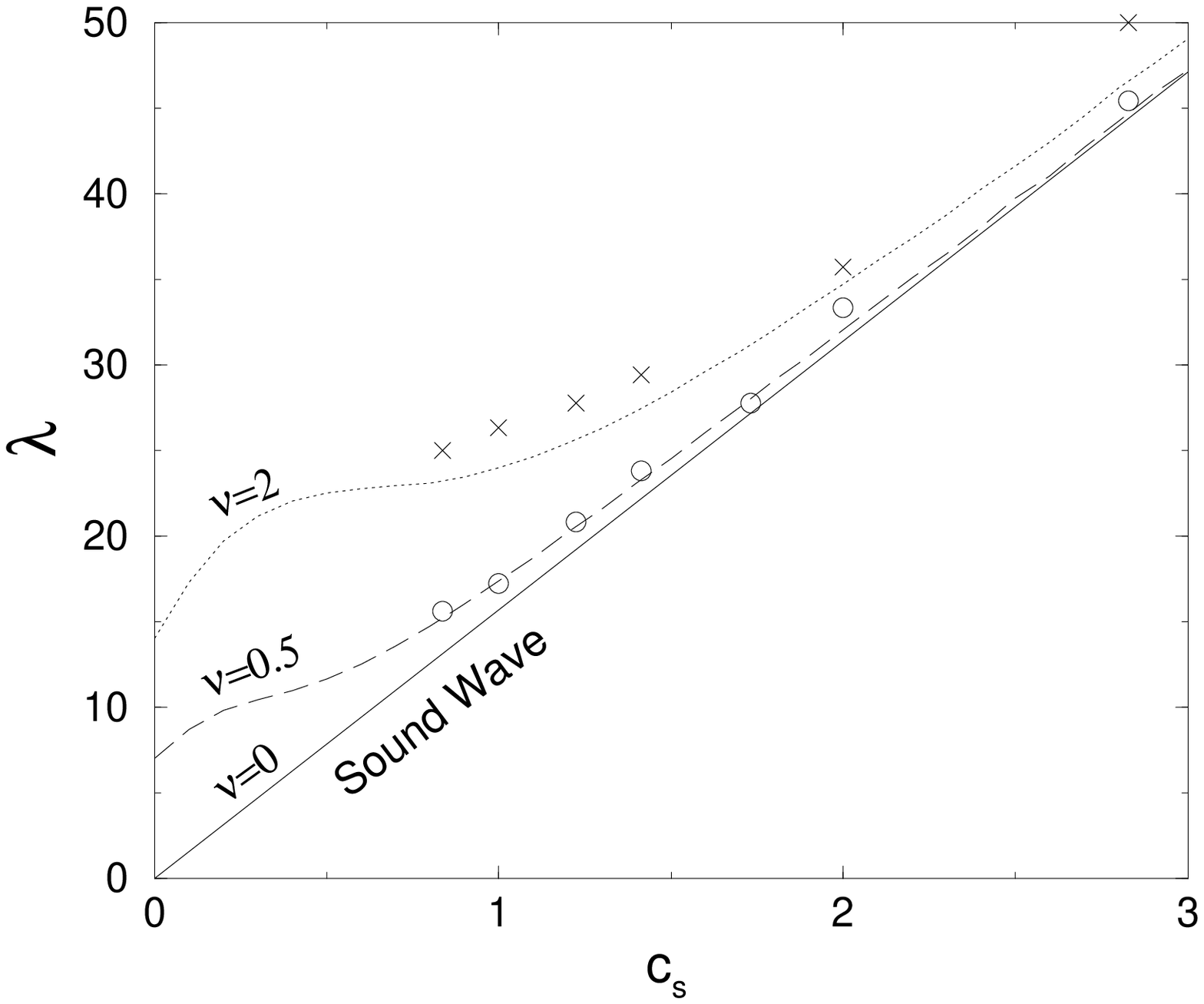}
\end{tabular} 
\end{center}
\caption{(a) Growth rate of stripes $\Gamma$ as a function of 
wavenumber $k$ ($\times$ simulations, - - - theory). 
(b) Critical forcing amplitude $a_c$, above which the system forms 
stripes, as a function of angular frequency $\omega$ for two 
different viscosities ($\times$ simulations with $\nu = 2$, $\circ$ 
simulations with $\nu = 0.5$, $\cdots$ theory with $\nu = 2$, 
 - - - theory with $\nu = 0.5$).
(c) Stripe wavelength $\lambda$ as a function of 
speed of sound $c_s$ for an angular frequency of $\omega=0.4$
and different values of viscosity $\nu$ 
(--- wavelength of sound).}
\label{grown}
\end{figure}

A typical curve of $\Gamma$ against $k$ is shown in Fig.~\ref{grown}(a). 
The peak represents the fastest growing wavenumber. The numerical
results, presented for comparison, were obtained by solving
Eqns.~(\ref{nss1})--(\ref{ce2}) for the parameters $\Delta t = 0.001$, 
$\omega = 0.2$, $\xi^0 = 32$, $\nu = 2$, $\theta = \frac{10}{3}$, 
$\rho_0 = 4$, $\Delta \phi = 0.001$, $a = 0.8$ and $k={2\pi}/{L}$ 
such that one stripe filled the entire system of length $L$.  
The initial condition was taken to be the trial solution.

Using (\ref{gama}) we may make two predictions. First, stripe 
formation only occurs when the amplitude of the force is above a 
certain critical value $a_c$ such that $\Gamma$ is positive for some
value of $k$.
To test this numerically we used a system with $L = 1500$, 
$\Delta t = 0.01$, $\xi^0 = 6$ and $\theta = 1$. Results are shown in
Fig.~\ref{grown}(b).
In each simulation we initially set the density to be $\rho_0 = 4$ 
plus a small random perturbation of amplitude $0.02$. Stripe formation 
was defined to occur when $\left<(\phi- \phi_0)^2 \right> > 0.01$ 
 and the system was observed for up to $20000$ time steps. 
Exact agreement between the analytic and numerical results is not
expected because the simulation size is finite and the value 
of $k$ is limited to integer multiples of $\frac{2\pi}{1500}$. 
Moreover the assumption that $\xi^0 >> \omega$ breaks down for 
large $\omega$.

Second, the wavelength of the stripes tends toward the wavelength 
of sound $\lambda_s = 2 \pi c_s/\omega$, where $c_s = \sqrt{\theta}$ for an isothermal gas, in the low
viscosity limit. This is clearly illustrated in Fig.~\ref{grown}(c), 
which show how stripe wavelength changes as a function of the speed of
sound for systems at $a_c$. The relation to $c_s$ is expected because
the evolution of the total density (\ref{nanb}) is a standing wave (i.e.
a sound wave)
which slowly increases in amplitude as the striped concentration
profile forms.

Thus we have shown that, within a continuum theory, differential
 forcing of the two components of a binary fluid can lead to stripe
 formation in the concentration profile.

We now take the first steps in investigating whether the same mechanism can hold in granular systems
 by solution of a two-dimensional, particle-based model which includes an
 excluded volume constraint. 
We consider two types of particles $A$ and $B$ which have equal diameter $d$, and equal mass $m$.
The particles undergo free streaming and a periodic force 
is applied to the $A$ particles.
\begin{eqnarray}
{\bf r}_i^{A,B} (t+\Delta t) &=& {\bf r}_i^{A,B} (t) + {\bf v}_i^{A,B} \Delta t,\\
{\bf v}_i^A (t+\Delta t) &=& {\bf v}_i^A (t) + {\bf a} \cos(\omega t) \Delta t.\label{vup}
\end{eqnarray}
Each particle is then identified as belonging to one of a grid of square 
cells with dimension $2d$.
Each particle is checked to see if it overlaps with particles 
in its own cell or the three other closest cells. If two overlapping particles are
 travelling towards each other
 then they collide elastically, thus conserving energy and momentum.
 The smaller the time-step 
$\Delta t$ the more rigorously the excluded volume condition is enforced.
The system is made isothermal by dividing it into slices of width $2d$
 perpendicular to the forcing direction and rescaling velocities in 
the center of mass frame of each slice to accord with equipartition
 of energy.

The system forms stripes of $A$ and $B$ particles perpendicular to the
forcing direction as in the continuum model. Moving to two dimensions
and including fluctuations does not qualitatively alter the stripe
formation mechanism.
However an
interesting difference which we do not yet have an explanation for is
that there is no threshold forcing amplitude for the stripe formation.
Instead the time taken to form stripes diverges as the forcing 
tends to zero. The wavelengths of the stripes corresponds to the
wavelength of sound to within $\sim 10\%$.

\begin{figure}
\begin{center}
\begin{tabular}{c}
\subfigure[t=0.25s]{\includegraphics[angle = 90, height = 4cm]{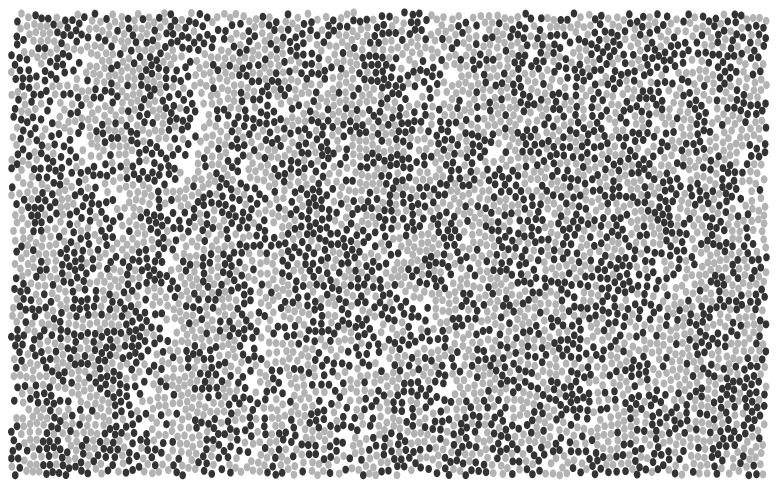}}
\subfigure[t=2.5s]{\includegraphics[angle = 90, height = 4cm]{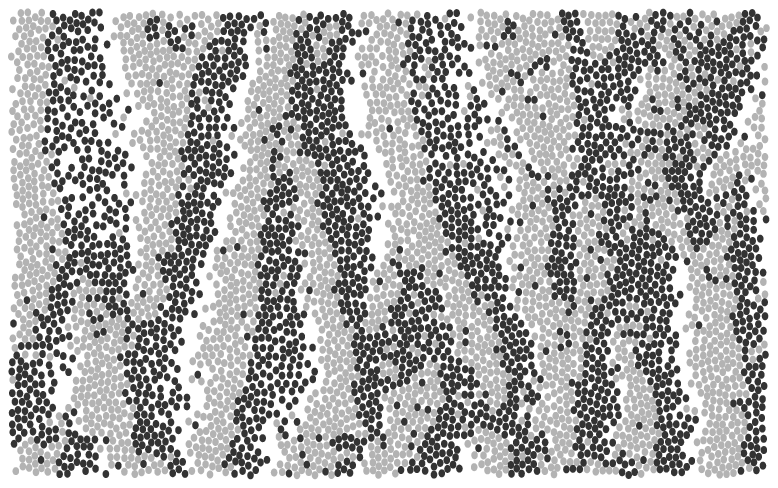}}
\subfigure[t=25s]{\includegraphics[angle = 90, height = 4cm]{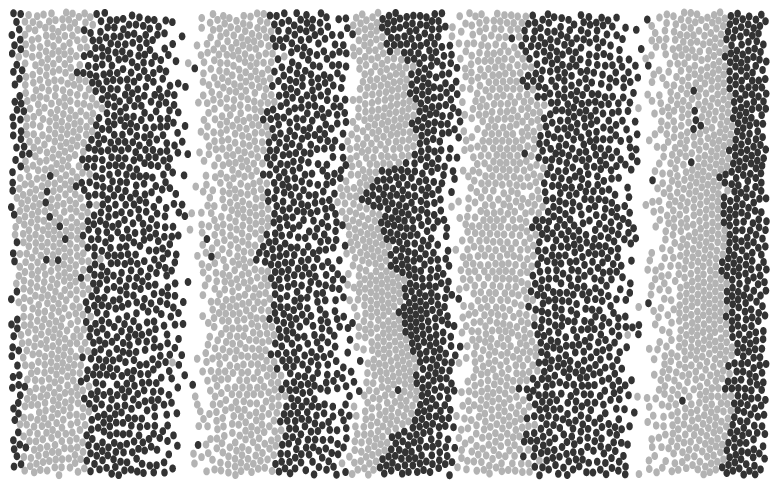}}
\end{tabular} 
\end{center}
\caption{Time evolution in seconds of a horizontally shaken granular
system. 
The different friction coefficients 
of the particle species cause the 
stripe formation. The parameters of the simulations are given in the text.} 
\label{ccg}
\end{figure}

Finally we discuss how  differential forcing of the two components 
of a granular mixture may be 
responsible for the experimental findings of Mullin {\it et al.} \cite{RM02}.
Mullin {\it et al.}  consider a flat tray which is oscillated in the 
horizontal plane at a frequency of $12Hz$.  
A mixture of poppy seeds and bronze spheres are placed on the tray. 
The particles are initially mixed and form 
an almost complete single particle layer. Forcing of individual 
particles arises principally from the frictional interaction with 
the base of the container. Mullin {\it et al.} observe that in time the two types 
of particles segregate into a striped pattern (very similar to that 
in Fig.~\ref{ccg}, which shows the results of the computer simulations 
described below).

We modelled this system using the excluded volume elastic scattering algorithm with the
update in the particle velocities given by
\begin{eqnarray}
\hspace{-0.3cm} {{\bf v}_i}^{A,B}(t\!+\!\Delta t) = {{\bf v}_i}^{A,B}(t)\! + \!\left({\bf a} \cos(\omega t)\! - \!\mu^{A,B} g \hat{{\bf v}}_i \right) \Delta t,
\end{eqnarray}
%
%
where
${\bf a}$ is the amplitude of the acceleration applied to both
species.  The final term represents the frictional interaction with
the container base: $g = 9.8 ms^{-2}$ is the acceleration due to gravity,
$\hat{{\bf v}}_i$ is the unit vector in the direction of 
the particle velocity, and $\mu^{A,B}$ are the 
coefficients of friction for poppy seeds and bronze spheres 
respectively. No thermostat is used. Energy from the
forcing is dissipated by friction.
Because the friction coefficients of the two types of particles 
are different they are differentially forced, although this difference at large $\bf a$
is like a square wave rather than the sinusoidal forcing used in Eqn. (\ref{vup}). 

 We make the approximation that the static and dynamic friction coefficients are equal and ignore
questions as to how much the spheres roll or slide. Experimental data for
$\mu^A$ and $\mu^B$ is not available and we use the physically realistic
values $\mu^A = 0.6$ 
and $\mu^B = 0.2$. We find that the qualitative behaviour 
is not dependent on the exact values.   

Fig.~\ref{ccg} is the result of a simulation showing the 
evolution in time of a shaken binary system which is initially mixed. 
The dimensions of the tray 
are $90mm$ by $180mm$ and there are $1500$ particles each with 
radius $r = 1.5mm$. The oscillation frequency was $4.8 Hz$ and the 
forcing amplitude $a = g$. Stripes form on a time scale 
very similar to that found in the experiments, which was $\sim 40s$ \cite {RM02}. Therefore it seems
reasonable that the differential acceleration of the two components can be
identified as at least partially contributing to the stripe
formation. However we caution that these simulations are not isothermal, hence direct comparison with
the isothermal model presented earlier is difficult. Furthermore the model assumes elastic collisions 
which may not be a good approximation for the real experiment.

To summarise, we have demonstrated analytically that if two 
isothermal ideal gases interacting via a
frictional force are subject to differential forcing
stripes form in the concentration profile above a critical forcing amplitude.
The stripe formation originates from
energy being pumped into standing wave oscillations, and consequently
the wavelength of the stripes is approximately that of the wavelength
of sound. These calculations  were carried out for a continuum
model. We then used a particle-based numerical model to provide
evidence that the same mechanism, differential forcing due to
differential friction with the base of the container, is likely to be
responsible for stripe formation in horizontally oscillated granular
mixtures. Further work will aim to use more realistic models of granular
systems to investigate the effect of inelastic collisions and
non-uniform temperature distributions on the pattern formation.

We thank M.R. Swift, T. Mullin and C. Denniston for helpful discussions.


\begin{references}
\bibitem{SM01} T. Shinbrot and F. Muzzio, Nature {\bf 410}: 251 (2001).
\bibitem{B97}  J. Bridgewater, Powder Technology {\bf 15}: 215 (1976).
\bibitem{W}    A. Wills, Mineral Process Technology, (Butterworth and Heinmann, 1997).
\bibitem{SS} P. S\'{a}nchez, M. R. Swift, and P. J. King, {\it preprint}.
\bibitem{RM02} P. M. Reis and T. Mullin, Phys. Rev. Lett. {\bf 89}: 244301 (2002).
\bibitem{M00}  T. Mullin, Phys. Rev. Lett. {\bf 84}: 4741 (2000).
\bibitem{RE03} P. M. Reis, G. Ehrhardt, A. Stephenson, and T. Mullin, Europhys. Lett. {\it in press} (2003).
\bibitem{AO58} S. Asakura and F Oosawa, J. Polym. Sci. {\bf 33}: 183 (1958).
\bibitem{MY99} A. Malevanets and J. M. Yeomans, Faraday Discuss. {\bf 112}: 237 (1999)
\end{references}
\end{document}